\documentclass[aps,prl,twocolumn,showpacs]{revtex4}
\usepackage{bm}
\usepackage{graphicx}
\usepackage{amsmath}
\usepackage{eufrak}
\usepackage{color}
\usepackage{upgreek}
\usepackage{subfigure}
\usepackage{hyperref}
\usepackage{ulem}
\newcommand{\nix}[1]{}

\begin{document}

\title{Photogalvanic probing of helical edge channels in 2D HgTe topological insulators}
\author{K.-M.\,Dantscher,$^1$  D.\,A.\,Kozlov,$^{2,3}$
M.\,T.\,Scherr,$^1$ S.\,Gebert,$^1$ J.\,B\"arenf\"anger,$^1$
M.\,V.\,Durnev,$^4$ S.\,A.\,Tarasenko,$^{4}$
V.\,V.\,Bel'kov,$^4$  N.\,N.\,Mikhailov,$^2$ 
S.\,A.\,Dvoretsky,$^2$ Z.\,D.\,Kvon,$^{2,3}$
D.\,Weiss,$^1$ and S.\,D.\,Ganichev$^1$}

\affiliation{$^1$Terahertz Center, University of
Regensburg, 93040 Regensburg, Germany}

\affiliation{$^2$A.V. Rzhanov Institute of
Semiconductor Physics, Novosibirsk 630090, Russia}
\affiliation{$^3$Novosibirsk State University,
Novosibirsk 630090, Russia}

\affiliation{$^4$Ioffe Institute,
194021 St.\,Petersburg, Russia}

\begin{abstract}
We report on the observation of a circular photogalvanic current
excited by terahertz (THz) laser radiation in helical edge channels of
HgTe-based 2D topological insulators (TIs).
The direction of the  photocurrent reverses by switching the
radiation polarization from right-handed to left-handed one
and, for  fixed photon helicity, is opposite for the opposite edges.
The photocurrent is detected in a wide range of gate 
voltages. With decreasing the  Fermi level below the conduction band 
bottom, the current emerges,  reaches a maximum, decreases, 
changes its sign close to the charge neutrality point 
(CNP), and again rises.
Conductance measured over a $7\,\upmu$m
distance at CNP approaches $2e^2/h$, the value characteristic for
ballistic transport in 2D TIs.
The data reveal that the 
photocurrent is caused by
photoionization of helical edge electrons to the conduction band. 
We discuss the microscopic model of this phenomenon and compare calculations with the experimental data.
\end{abstract}

\pacs{}
\maketitle

The quantum spin Hall (QSH) effect occurs in two-dimensional 
topological insulators and rests on the existence of conducting 
helical edge  states while the bulk of the two-dimensional system 
is insulating~\cite{Kane2005, Bernevig2006, Hasan2010, Zhang2011}. 
In contrast to the quantum Hall effect, the formation of these 
edge states requires no magnetic field $\bm{B}$: they stem from
the band inversion caused by strong spin-orbit interaction and are, 
due to the absence of $\bm{B}$, topologically protected by time reversal 
symmetry. Given that the spin-up and spin-down electrons 
propagate along an edge in the opposite  directions, 
i.e., the spin projection  is locked to the $\bm{k}$-vector, 
the edge channels are helical in nature. The first experimental evidence for 
the QSH effect was obtained in HgTe quantum wells (QWs)~\cite{Konig2007} 
by observing a resistance plateau around $h/(2 e^2)$ in the longitudinal 
resistance of a mesoscopic Hall bar. Here, $h$ is Planck's constant and 
$e$ is the elementary charge. This observation was further confirmed by 
non-local experiments both in the ballistic~\cite{Roth2009} and diffusive~\cite{GusevPRL2011} 
transport regime.  Conducting edge channels were  later probed by scanning 
squid microscopy~\cite{NowackNatureMat2013}, scanning gate microscopy~\cite{KonigPRX2013}, 
microwave impedance microscopy~\cite{MaNatureComm2015}, and by analyzing the 
spatial distribution of super-currents~\cite{HartNaturePhys2014}. The spin polarization 
of the edge states was investigated so far by electrical means only: by detecting
 the spin to charge conversion in devices utilizing the inverse spin Hall 
effect~\cite{BruneNaturePhysics2012} or with ferromagnetic contacts~\cite{KononovJETPLett2015}. 

Here, we use circularly polarized THz 
radiation to excite selectively spin-up and spin-down 
electrons circling clockwise and counter-clockwise around the sample edges. 
We show that the selective excitation causes an imbalance in the electron 
distribution between positive and negative wave vectors. This is 
 probed  as the associated photocurrent, which reverses its
direction upon switching the helicity of the radiation 
polarization.

\begin{figure}[t]
\includegraphics[width=\linewidth]{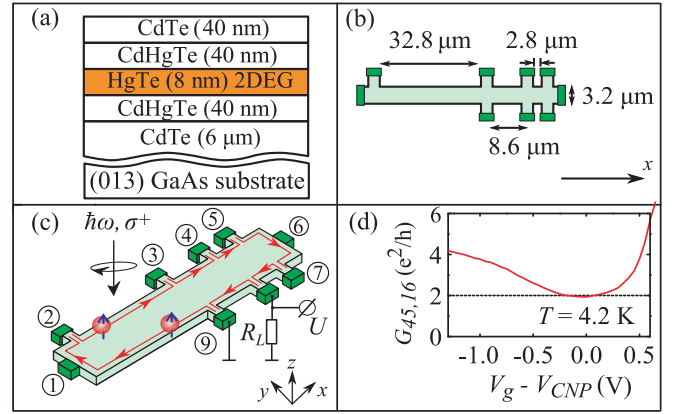}
\caption{
Cross section (a) and geometry (b) of the studied structures. 
(c) Experimental setup of the photocurrent measurement.
Arrows along edges illustrate spin polarized 
current excited by circularly polarized radiation.
(d) Conductance measured for sample \#\,1, showing 
ballistic transport in the vicinity of CNP.
}
\label{fig_1}
\end{figure}

The experiments have been carried out on high-mobility 
Hg$_{0.3}$Cd$_{0.7}$Te/HgTe/Hg$_{0.3}$Cd$_{0.7}$Te single QW structures
with a well width $L_{w}=8$~nm,
having the inverted band ordering~\cite{Zhang2011,Konig2007}. 
The structures were grown by molecular beam epitaxy on  
(013)-oriented GaAs substrates~\cite{Kvon2008,ellipticitydetector2}, 
the corresponding layer sequence is  shown in Fig.~\ref{fig_1}(a).
Several 
samples have been prepared  from the same wafer. The typical Hall bar design, 
dimensions, and the ohmic contacts positions on the device  are shown in 
Figs.~\ref{fig_1}(b)  and (c). This geometry allows us to study both, 
magneto-transport and photocurrents induced by THz  radiation along the 
circumference of a sample.

The devices 
have been patterned by means of 
photo\-litho\-graphy and reactive ion 
inductively coupled plasma etching with hydrogen. In order to vary the 
Fermi level position 
 the devices are equipped with semitransparent Ti(15 nm)/Au(5 nm) 
gates on top of a $200$\,nm SiO$_{2}$ 
layer
 grown by chemical vapour deposition. All samples have been characterized by 
transport measurements 
with currents in the range of $10-100\,$nA. The mobility measured 
at $T=4.2$~K
is about $10^5$\,cm$^2$/(V$\,$s) at the QW
 carrier density of 3$\cdot$10$^{11}$\,cm$^{-2}$ at zero gate voltage.

Figure~\ref{fig_1}(d) shows the 4-terminal conductance at 4.2\,K measured for the voltage drop
between the contacts~4 and 5 and the current flowing between the contacts~1 and 6.
It  demonstrates the conductance quantization close to $2e^2/h$, i.e., the system is tuned into the QSH regime.
The conductance  around $2e^2/h$  is only detected for the closest contact pairs 
4-5 and 7-8 (the contact spacing is 2.8\,$\mu$m and the edge length is about 7\,$\mu$m).
For contact pairs with larger separation, e.g. for the contact pair 3-4, the conductance ranges between $2e^2/h$ and $e^2/h$. For different sample cool-downs, 
the CNP
can occur at different gate voltages. 
This is caused by the to cool-down dependent charge trapping in the insulator. 
While we find the CNP sometimes shifted, the overall behavior of the signals remains unchanged. 
To compare the measurements taken at different sample 
cool-downs we plot the data as a 
function of the normalized gate voltage $ V_g - V_{\rm CNP}$ with $V_{\rm CNP}$ being
 the gate voltage of CNP.

We excite photocurrents by applying circularly polarized THz radiation 
of a continues wave (\textit{cw}) molecular laser~\cite{DMSPRL09,Drexler13} under normal incidence. 
Two radiation frequencies $f$ were chosen: (i)  $f = 2.54$ THz with the photon energy $\hbar\omega = 10.4$\,meV and (ii) $f = 1.62$ THz with $\hbar\omega = 6.7$\,meV. 
The laser beam with the power $P \approx 10$\,mW and an 
almost Gaussian profile,
measured by a pyroelectric camera~\cite{Ziemann2000}, 
is focused onto a spot of about $1.5$\,mm diameter, thus 
illuminating the whole sample. The radiation intensity 
$I \approx 0.6$\,W/cm$^2$ is modulated by an optical chopper at the frequency 600~Hz.
To create right-handed ($\sigma^+$) and left-handed 
($\sigma^-$)  circularly polarized radiation  $\lambda/4$ plates are used. 
We study the photosignal generated in the sample by measuring the voltage drop $U$ 
across a load resistance $R_L$, see Fig.~\ref{fig_1}(c), by using two 
electrical measurement configurations and standard lock-in technique:   
(i)  $R_L = 50\,\Omega$ with $R_L\ll R_s$,
where $R_s$ is the sample resistance or (ii)  $R_L \gg R_s$. 
In the former case the photocurrent is given by $J = U/R_L$.
All experiments are performed at liquid helium temperature.

\begin{figure}[t]
\includegraphics[width=\linewidth]{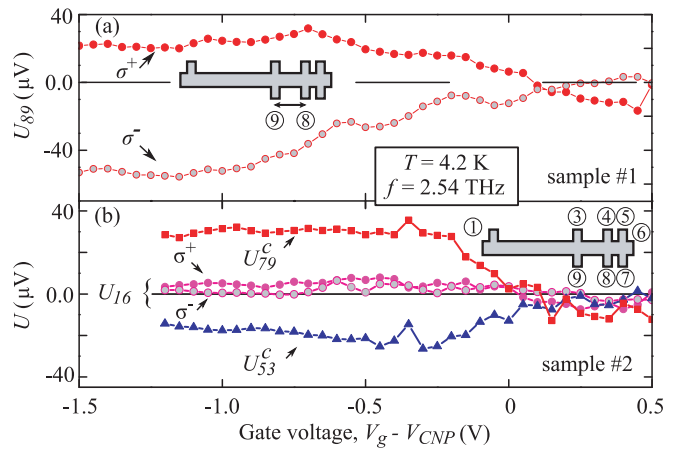}
\caption{(a) Photovoltage excited by right-handed ($\sigma^+$, full circles) and left-handed  ($\sigma^-$, open circles) circularly polarized radiation and measured between contacts~8 and 9 (see inset)
as a function of  the gate voltage. (b) Photon helicity sensitive photoresponse, 
$U^c= [U(\sigma^+) - U(\sigma^-)]/2$,
obtained for sample \#2 for two opposite edges
as a function of gate voltage. 
Full and open circles show the photoresponse $U_{16}$
excited by $\sigma^+$ and $\sigma^-$ radiation and measured 
over the whole sample.
}
\label{fig_2}
\end{figure}

When illuminating an unbiased devices with circularly polarized radiation we detect
a photovoltage $U$ between any pair of contacts along the same edge.  
The dependence of the photovoltage $U_{89}$, measured between the contacts~8 
and~9 at one edge of the sample, on the gate voltage is shown in Fig.~\ref{fig_2}(a). 
The central observation is that the polarity of $U_{89}$ changes upon changing 
the polarization from $\sigma^+$ to $\sigma^-$. 
Furthermore, the helicity dependent signals defined 
as $U^c= [U(\sigma^+) - U(\sigma^-)]/2$, 
plotted in Fig.~\ref{fig_2}(b), show consistently different polarity for contacts pairs 
on the opposite sides of the sample. 
This indicates that the photoresponse stems from a 
photocurrent flowing along the edges of the sample. 
The sense of circulation of the photocurrent depends on the photon helicity.
Measurement of $U_{16}$ across the sample,  
between the contacts~1 and 6, shown in Fig.~\ref{fig_2}(b), 
confirms this scenario:  the signals are 
vanishingly small, which is 
ascribed to the compensation
of the counter propagating currents generated along the opposite edges. 
Interestingly, the signs of 
the voltages caused by the \textit{edge} photosignal  reverse at the gate voltage
between~0 and 0.5 V heralding that the sense of circulation
of the edge current changes as a function of the gate voltage, see  Fig.~\ref{fig_2}(b).

\begin{figure}[t]
\includegraphics[width=\linewidth]{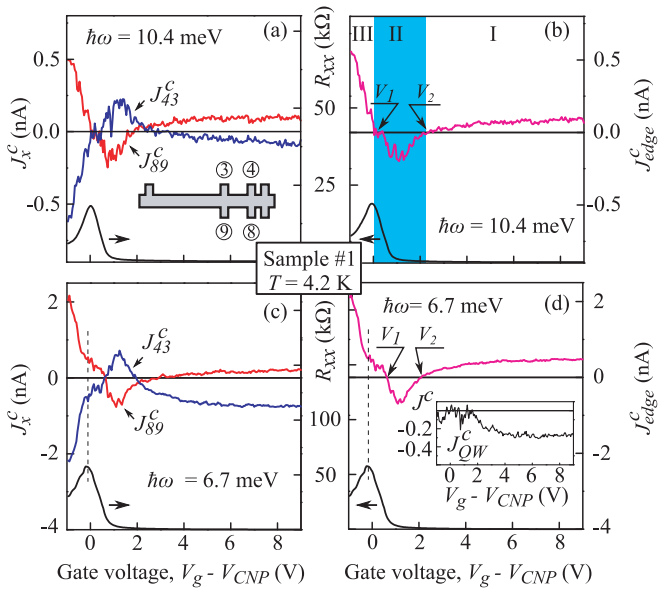}
\caption{
(a) and (c): Circular photocurrents $J^c_x= [J(\sigma^+) - J(\sigma^-)]/2$ obtained for $R_L \ll R_s$
for two contact pairs at the opposite edges, see inset in panel (a). Black solid
curves show the sample resistance measured between the contacts~3 and~4. (b) and (d)  Edge
and QW bulk (inset)  contributions.
}
\label{fig_3}
\end{figure}

The measured circular photocurrent $J^c_x= [J(\sigma^+) - J(\sigma^-)]/2$, 
obtained for $R_L \ll R_s$, is displayed in Figs.~\ref{fig_3}(a) 
and~(c) and exhibits similar behavior. 
The data are shown for two THz frequencies corresponding to 
$\hbar \omega =10.4$\,meV (panel a) and $\hbar \omega =6.7$\,meV (panel c), 
i.e., the photon energies smaller than the bulk gap. 
 The fact that \textit{edge} currents at the opposite sides of the sample have, in contrast to \textit{bulk} currents, opposite 
polarities allows us to refine the edge current contribution $J^c_{\rm edge}$ 
by subtracting the currents measured at the opposite sides, $J^c_{\rm edge} = (J^c_{43}-J^c_{89})/2$.
The corresponding data 
are shown in Figs.~\ref{fig_3}(b) and~(d).
By summing up the currents, $J^c_{\rm QW} = (J^c_{43}+J^c_{89})/2$,  we obtain the bulk contribution 
which is shown in the inset of Fig.~\ref{fig_3}(d).
Remarkably, the sign of the edge current $J_{\rm edge}^{c}$ changes twice:
 at a gate voltage close to the CNP and at $V_g - V_{\rm CNP} \approx 2$\,V, while 
the bulk contribution $J^c_{\rm QW}$ is noticeable
 only for larger gate voltages. The central result so far
is that we can selectively excite left- and right-moving edge currents by means of the 
helicity of circularly polarized radiation, in accordance 
with the helical nature of these states. Below we resort to a microscopic model to 
understand  the origin of the photocurrents in the different gate voltage
regions, marked by~I to~III in Fig.~\ref{fig_3}(b). Region~II corresponds to the gate 
voltages between the nodes of the photocurrent,
whereas regions~I and~III correspond to 
gate voltages above and below the borders of the region~II. 
Note that the border between region~II and~III is close to the CNP.

\begin{figure}[t]
\centering
\includegraphics*[width=\linewidth]{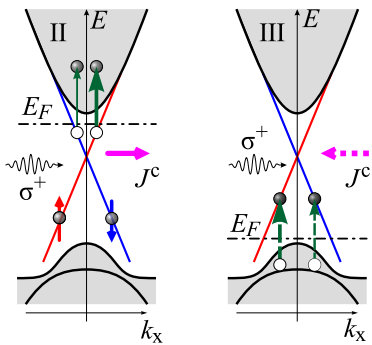}
\caption[]
{ 
Schematic picture of optical transitions for right-handed circularly polarized radiation and the
photocurrent formation for two positions of the Fermi level $E_{\rm F}$ in the bulk gap. $k_x$
is the wave vector along the sample edge.  Region~II:  $E_{\rm F}$ is below the conduction-band 
bottom $E_c$ and {$E_{\rm c} - E_{\rm F} < \hbar \omega$}. 
Region~III: $E_{\rm F}$ is close to the valence band top.
}
\label{fig_ti1}
\end{figure}

In region~I, i.e., at  $V_g - V_{CNP} > 2$~V, the Fermi level 
lies in the conduction band, as it follows from transport measurement, 
see the resistance $R_{xx}$ in Fig.~\ref{fig_3}(b).
In this region, both bulk and  edge photocurrents are formed
 by conduction-band carriers. The reduced symmetry at the sample edges results in an asymmetric scattering of carriers  causing an edge current with the direction determined by the helicity of the THz field. This mechanism has been explored before in graphene and other 2D materials~\cite{edgePRL2011,GlazovGanichev} and is also active here.

If $E_{\rm F}$ is in the bulk gap the above mechanism is no longer effective 
and the presence of topological edge states needs to be taken into account to explain the current formation.
Figure~\ref{fig_ti1} illustrates the linear dispersion of helical edge states between the schematic valence and conduction band. The  branches of the dispersion are spin polarized as marked by arrows in panel (a). 
The edge states with the positive velocity along $x$ direction
are formed mainly from $|E1, +1/2 \rangle$ and $|H1, +3/2 \rangle$ 
subbands and have pseudospin $s=+1/2$ (spin-up branch)~\cite{Durnev2016,Tarasenko2015}. 
Counter propagating electrons
have $s=-1/2$ (spin-down branch) and these edge states are  
mainly formed from $|E1, -1/2 \rangle$ and $|H1, -3/2 \rangle$ subbands.

The Fermi level crosses the bottom of the conduction band 
and enters the bulk gap at  gate voltage about 2\,V. This corresponds to region~II displayed in Fig.~\ref{fig_ti1}(a).
While the bulk photocurrent ceases, see the inset of  Fig.~\ref{fig_3}(d), the edge photocurrent changes polarity, rises, and  exhibits a maximum at  
$V_g-V_{CNP} \approx 1$~V. 
From the facts that in region II 
$E_{\rm F}$
is in the bulk gap and the photon energy is smaller than the band gap we conclude that the observed current is caused by the excitation of electrons from helical edge states into higher lying states. Different optical transitions are conceivable including those between the edge and bulk states, direct optical transitions between the spin-up and spin-down branches of the 
linear edge-state spectrum, and
indirect 
transitions.
Comparing the efficiencies of all the above processes~\cite{footnotehgte1}
we attribute the circular edge photocurrents observed in the region~II
to the excitation of electrons from helical edge states to bulk conduction-band states
(``photoionization'' of the edge channels)~\cite{Kaladzhyan2015,Magarill2016}.

The generation of circular photocurrents via the edge-to-bulk excitation is
schematically shown in Fig.~\ref{fig_ti1}(a). It involves similar physical concepts as used for describing photocurrents
due to inter-subband  transitions in semiconductor QWs~ \cite{PRB03inv}.
Vertical arrows in Fig.~\ref{fig_ti1} represent the photoionization of helical states, i.e., the depopulation of Dirac states and population of excited bulk states. The arrows end above the conduction band bottom illustrating the transitions to bulk states with nonzero transverse wave vectors $k_y$.
For the excitation with circularly polarized radiation, e.g., $\sigma^+$,
the probability of optical transitions from the 
 states with $s=1/2$ is larger than those involving the 
states with $s=-1/2$. This is illustrated in
Fig.~\ref{fig_ti1}(a) by arrows of different thicknesses.
The resulting imbalance of the edge-state populations leads  to a net electric current $J^c$ (horizontal solid arrow). 
For incident  radiation of the opposite helicity $\sigma^-$, the optical transitions
from the spin-down states  dominates and the photocurrent reverses its direction.

The described above selection rules for the edge-to-band optical transitions can be written as
\begin{equation}
\frac{g_{+1/2}(k_x) - g_{-1/2}(-k_x)}{g_{+1/2}(k_x) + g_{-1/2}(-k_x)} = K P_{circ}  \,,
\end{equation}
where $g_{\pm 1/2}$ are the probabilities of optical transitions from the initial state with $s = \pm 1/2$, $K$ is the coefficient describing the rigidity of the selection rules,
and $P_{circ}$ is the radiation helicity. Our calculations show that  $K$, determined by the band structure parameters,  depends only weakly on  $k_x$ and the photon energy $\hbar\omega$. In a simple 4-subband model with inversion center, the coefficient $K$ at $k_x=0$
is given  by $2 B D/(B^2 + D^2)$, Ref.~\cite{Kaladzhyan2015},
where $B$ and $D$ are the parameters of the Bernevig-Hughes-Zhang (BHZ) Hamiltonian~\cite{Bernevig2006}.
Note that the photoexcited carriers in the conduction band also contribute to the photocurrent but this contribution is small due to fast momentum relaxation $\tau_p$ of bulk carriers.

\begin{figure}[htpb]
\includegraphics[width=\linewidth]{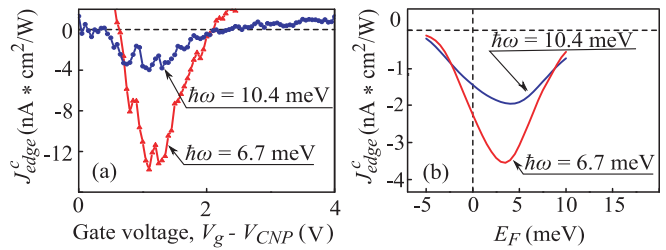}
\caption{\label{fig:fig1} {
(a) Circular edge photocurrents measured between the nodes
of the gate dependencies.
(b) Calculated edge photocurrents as a function of the Fermi level position $E_{\rm F}$ counted from the Dirac point.}
}
\end{figure}

The edge photocurrent based on the above photoionization picture requires that the initial states of the optical 
transitions are occupied and the final ones are empty and that the energy conservation holds. The corresponding energetic window is given by $E_c - \hbar \omega < E_F < E_c + \hbar\omega$. 
In real structures, the window is broadened due to inhomogeneities of the structures. Nevertheless, the range of Fermi levels and, respectively, the range of gate voltages
for which this photocurrent is efficiently generated increases with increasing the photon energy $\hbar \omega$. Exactly such a behavior is observed in  experiment, see Figs.~\ref{fig_3}(b),(d) and Fig.~\ref{fig:fig1}(a).

Calculations within the relaxation time approximation yield the following expression for the edge photocurrent
\begin{align}
j_x  = - \frac{e K P_{circ}}{2 \pi \hbar} \int \tau_p(\varepsilon) g_{{\rm tot}} (\varepsilon) d \varepsilon \,, \label{j_edge}
\end{align}
where $e$ is the electron charge,
$\tau_p$ is the momentum relaxation time determined  by backscattering processes for edge carriers, $ g_{{\rm tot}} (\varepsilon) = g_{+1/2} + g_{-1/2}$ is the total rate of photoionization of  electrons with the energy $\varepsilon$ by circularly polarized radiation. 
Here, we assume that the spin relaxation of bulk carriers  is fast and, therefore, photoexcited carriers get unpolarized before they are trapped back on helical edge states. This means that the depopulation of edge states determines the photocurrent.

Figure~\ref{fig:fig1} compares the calculated photocurrents with the measured ones for the photon energies $\hbar\omega = 6.7$ and $10.4$~meV, 
magnified view of the data shown in Figure~\ref{fig_3}(b) and (d).
In the calculations, 
displayed in Fig.~\ref{fig:fig1}(b), we assume that $\tau_p(\varepsilon)$ is a step-like function: $\tau_p = 20$~ps for energies in the band gap,  which is estimated from the length of ballistic transport edge $\sim 7$~$\mu$m in our devices and the edge state velocity $\sim 10^7$~cm/s, and $\tau_p = 0.3$~ps for energies 
above the conduction band bottom, which is estimated from the bulk mobility. It is seen that with a decrease of the radiation frequency the photocurrent peak narrows. Figure~\ref{fig:fig1}(b) shows that the theory describes  the dependence of the photocurrent on the gate voltage  and the radiation frequency quite well. Moreover, the calculated magnitude of the current is close to the experimental values.

Finally, we discuss region III. This region corresponds to negative gate voltages where $E_{\rm F}$ is shifted towards the valence band which in  8-nm-wide QWs has a quite complex structure. Similarly to region II, we attribute the observed photocurrent, which
has the opposite sign, to optical transitions from the valence-band states to the helical edge states, 
see Fig.~\ref{fig_ti1}(b). This mechanism of photocurrent formation is essentially the same as in region II. 
The only difference is that the initial states are now in the valence band and the final ones 
are edge states. The question, though, is  why the photocurrent direction reverses. 
To describe the current direction for a given helicity correctly one needs to assume that the selection rules for optical transitions are reversed, i.e., the transitions to the $s=+1/2$ edge states occur now dominantly for right-handed circularly polarized radiation. This, however, is at odds with the simple BHZ model Hamiltonian. In fact, this is not surprising. The valence-band structure of 8-nm-wide HgTe QWs is known to be strongly affected by the closely lying excited $|H2\rangle$ subband not included in the BHZ model, which results in a non-monotonic dispersion of the hole states and the formation of side maxima~\cite{Minkov2016}. The current reversal may be also related to a strong energy dependence of the momentum relaxation time. A calculation of the optical transitions and the photocurrent requires the detailed knowledge of the real band structure and optical transitions beyond the BHZ model and will be done elsewhere.

To summarize, our results demonstrate that excitation of  HgTe-based 2D TIs by circularly
polarized THz radiation results in a spin polarized \textit{dc} 
electric current flowing in helical edge channels. This observation provides a novel access for probing spin transport in TIs.

\begin{acknowledgments}  We thank G.V. Budkin for helpful discussions. The support by the DFG (SPP 1666), Elite Network of Bavaria (K-NW-2013-247), and
the Volkswagen Stiftung Program, and the RFBR (projects 15-02-06344, 16-02-01037, 16-12-10041, 16-32-00540, and 16-32-60175) is gratefully acknowledged.
\end{acknowledgments}


\begin{thebibliography}{99}
\bibitem{footnotehgte1}
Direct optical transitions between the spin-up and spin-down branches have a low probability.
Moreover, such a contribution would have a maximum for the Fermi energy lying at the Dirac point and would vanish for large negative
or positive bias voltages, which is not the case. Photocurrents caused by the Drude-like free-carrier absorption require additional scattering and $\bm k$-dependent velocity and, therefore, seem to be weak. 
It can, however, provide a  contribution if one takes the energy dependence of the back scattering time into account.
\end{thebibliography}
\end{document}